\documentclass{elsart}

\usepackage[totalwidth=14.0cm, totalheight=22.8cm]{geometry}
\usepackage{natbib,graphicx}

\newcommand{\oasis}{{\texttt {OASIS}}}
\newcommand{\sauron}{{\texttt {SAURON}}}
\newcommand{\gmos}{{\texttt {GMOS}}}
\newcommand{\osiris}{{\texttt {OSIRIS}}}
\newcommand{\sinfoni}{{\texttt {SINFONI}}}
\newcommand{\nifs}{{\texttt {NIFS}}}
\newcommand{\kms}{$\mbox{km s}^{-1}$}
\newcommand{\arcsec}{\hbox{$^{\prime\prime}$}}  
\newcommand{\farcsec}{\hbox{$.\!\!^{\prime\prime}$}}  
\newcommand{\lsim}{\mathrel{\hbox{\rlap{\hbox{\lower4pt\hbox{$\sim$}}}\hbox{$<$}}}}
\newcommand{\gsim}{\mathrel{\hbox{\rlap{\hbox{\lower4pt\hbox{$\sim$}}}\hbox{$>$}}}}
\begin{document}
\runauthor{McDermid et al.}
\begin{frontmatter}
\title{Stellar Kinematics and Populations of Early-Type Galaxies with the \sauron\/ and \oasis\/ Integral-Field Spectrographs}
\author[Leiden]{Richard M. McDermid,\thanksref{email}}
\author[Lyon]{Roland Bacon,}
\author[STECF]{Harald Kuntschner,}
\author[Lyon]{Eric Emsellem,}
\author[Berkeley]{Kristen L. Shapiro,}
\author[Oxford]{Martin Bureau,}
\author[Leiden]{Michele Cappellari,}
\author[Oxford]{Roger L. Davies,}
\author[Leiden]{Jes\'us Falc\'on-Barroso,}
\author[Oxford]{Davor Krajnovi\'c,}
\author[Groningen]{Reynier F. Peletier,}
\author[Oxford]{Marc Sarzi,}
\author[Leiden]{Tim de Zeeuw}

\address[Leiden]{Leiden Observatory, Leiden, The Netherlands}
\address[Lyon]{Centre de Recherche Astronomique de Lyon, Lyon, France}
\address[STECF]{Space Telescope European Coordinating Facility, Garching, Germany}
\address[Berkeley]{University of California, Berkeley, USA}
\address[Oxford]{University of Oxford, Oxford, UK}
\address[Groningen]{Kapteyn Astronomical Institute, Groningen, The Netherlands}
\thanks[email]{mcdermid@strw.leidenuniv.nl}

\begin{abstract} We summarise the results and achievements of integral-field spectroscopy of early-type galaxies, observed as part of a survey using both the \sauron\/ and \oasis\/ spectrographs. From the perspective of integral-field spectroscopy, these otherwise smooth and featureless objects show a wealth of structure, both in their stellar kinematics and populations. We focus on the stellar content, and examine properties on both kiloparsec scales with \sauron, and scales of 100's of parsecs with \oasis. These complementary studies reveal two types of kinematically distinct components (KDCs), differing primarily in their intrinsic sizes. In previous studies, KDCs and their host galaxies have generally been found to be unremarkable in other aspects. We show that large KDCs, typical of the well-studied cases, indeed show little or no age differences with their host galaxy. The KDCs detected with the higher spatial-resolution of \oasis\/ are intrinsically smaller and include, in contrast, a significant fraction of young stars. We speculate on the relationship between KDCs and their host galaxies, and the implications for young populations in early-type galaxies.\looseness-2

\end{abstract}


\end{frontmatter}
%
%
\section{Early-Type Galaxies}

For a long time, the conventional view of early-type galaxies was that they were simple, single-component objects which were either almost spherical and hardly rotating, or flattened systems whose shape could be simply accounted for by rotation around a single axis. As telescopes became larger, and instruments and detectors became more efficient, it was possible to obtain a more detailed picture of the stellar motions of these systems. It became clear that many of these objects rotated too slowly to account for their apparent flattening: extra pressure support, in the form of an anisotropic velocity ellipsoid, was required to explain their shape \citep{bertola75,illingworth77,binney78}. Overall, early-type galaxies seemed to span a broad range of dynamical states, from rotating disk-like systems to dynamically hot, pressure-supported spheroids \citep[and references therein]{binney82,dezeeuw91}.

Furthermore, some objects were found to exhibit peculiar stellar kinematics, rotating around both the apparent short- and long-axes, for example, or showing a reversal in the rotation direction close to the galaxy center. Such galaxies were said to contain kinematically decoupled components (KDCs), indicating that there existed a sub-component of the galaxy which did not share the same orbital distribution as the rest of the object. The formation of such a sub-component can occur either by the accretion of an external system \citep[e.g.][]{kormendy84, balcells90}, or by the formation of stars from accreted external material \citep[e.g.][]{hernquist91,weil93}. Both scenarios have different implications for our theoretical understanding of galaxy formation.

Our understanding of the morphology of early-type galaxies has undergone a similar evolution. In particular, elliptical galaxies, as their name implies, were once considered to have apparent shapes described by a simple ellipse. With the advent of CCD detectors, improved photometry showed that elliptical galaxy isophotes exhibit small but measurable deviations from a simple elliptical shape. In general, elliptical galaxies can be divided into `boxy' and `disky' objects, quantified by a Fourier expansion around the best-fitting ellipse \citep[e.g.][]{bender88}. This property was found to correlate with other quantities \citep[such as rotational support, and even X-ray luminosity:][]{bender89}, indicating a close link between the morphology and evolution of these objects. Deep imaging studies have also shown that at very large radii, early-type galaxies can show faint asymmetric or `shell'-like features \citep[e.g.][]{schweizer90}. Such `fine structure' features are generally interpreted as evidence for interactions of some kind, perhaps remnants of the galaxy's formation.

Colour gradients were also observed, suggesting that early-type galaxies were not composed of a single population of stars, but were in fact composite populations, albeit rather smoothly varying \citep[e.g.][]{peletier90}. With further improvement in observational data, it became possible to study the individual elemental absorption lines from these stars, allowing the effects of increasing age and metallicity (which have a degenerate influence on broad-band colours) to be decoupled. By comparing the strength of various absorption lines (via equivalent width measurements, known as line-strength indices) in galaxies and in stellar population models, it was found that early-type galaxies are generally old, evolved systems, but that a significant fraction still appear to contain young stellar populations.\looseness-2

The spread in the apparent (luminosity-weighted) age of early-type galaxies is approximately 2--15~Gyr \citep[e.g.][]{trager00}. The majority of studies on significant samples of objects have been limited to aperture measurements, thus ignoring information on the spatial distribution of different populations within a given object. Moreover, because young populations are generally brighter than old populations, a small fraction of young stars {\em by mass} can result in a dramatic reduction in the integrated {\em luminosity-weighted} age of the system. This spread in age is therefore indicative of a spread in recent star-formation histories between these galaxies. How are the young stars distributed in these galaxies? And what physical conditions lead to the presence of young stars in some galaxies and not others?\looseness-2

The importance of early-type galaxies in the context of galaxy formation and evolution is paramount. The evolved nature of these objects endow them with a wealth of so-called `fossil' evidence of the processes of galaxy evolution. In the nearby universe, we can study these objects in far greater detail than at higher redshifts, allowing us to uncover clues from which we can re-trace the steps along the evolutionary path of the galaxy. From the above empirical evidence, early-type galaxies in the nearby universe can show signs for relatively recent formation: disturbed morphologies, kinematic sub-components, and young stellar populations. However, studies at high redshift indicate the existence of a population of objects which have properties similar to those of `conventional' early-type galaxies: smooth, spheroidal morphologies, dynamically hot, with evolved populations and little or no ongoing star-formation \citep[e.g.][]{franx03,chapman04,treu05}. Consolidating the various observational evidence concerning the origin and evolution of early-type galaxies remains a key challenge for galaxy formation theories.\looseness-2
\newpage
\section{The \sauron\/ Survey}

The development of the study of early-type galaxies is illustrative of how many advancements in astronomy are linked to advancements in instrumentation (not just telescope aperture). Until recently, one of the limiting factors in our understanding of these extended objects was having information only along one or a few axes from a long-slit spectrograph. Integral-field spectroscopy provides the next leap forwards in our perspective on these galaxies. At the forefront of this development is the \sauron\/ survey. In response to the aforementioned scientific questions, we have undertaken a survey of 72 representative early-type galaxies (24 each of ellipticals (Es), lenticulars (S0s) and Sa spiral bulges) using our custom-built panoramic integral-field spectrograph, \sauron, mounted on the William Herschel Telescope (WHT), La Palma \citep[Paper I]{bacon01}. With this survey, we map both the stellar kinematics and line-strengths within the  4800-5300\AA\/ wavelength region, and extract properties of the ionised gas where present. A review of the project aims is given in \citet[Paper II]{dezeeuw02}, and maps of the stellar kinematics, ionised gas properties and absorption line-strengths of the 48 E and S0 galaxies are given in \citet[Paper III]{emsellem04}, \citet[Paper V]{sarzi05} and \citet[Paper VI]{kuntschner05} respectively (for a summary of Paper V, see the contribution of Falc\'on-Barroso et al. in these proceedings).

\subsection{Stellar Kinematics with \sauron}
\label{sec:sau_zoo}

\begin{figure}
\begin{center}
\includegraphics[width=14.5cm, angle=0]{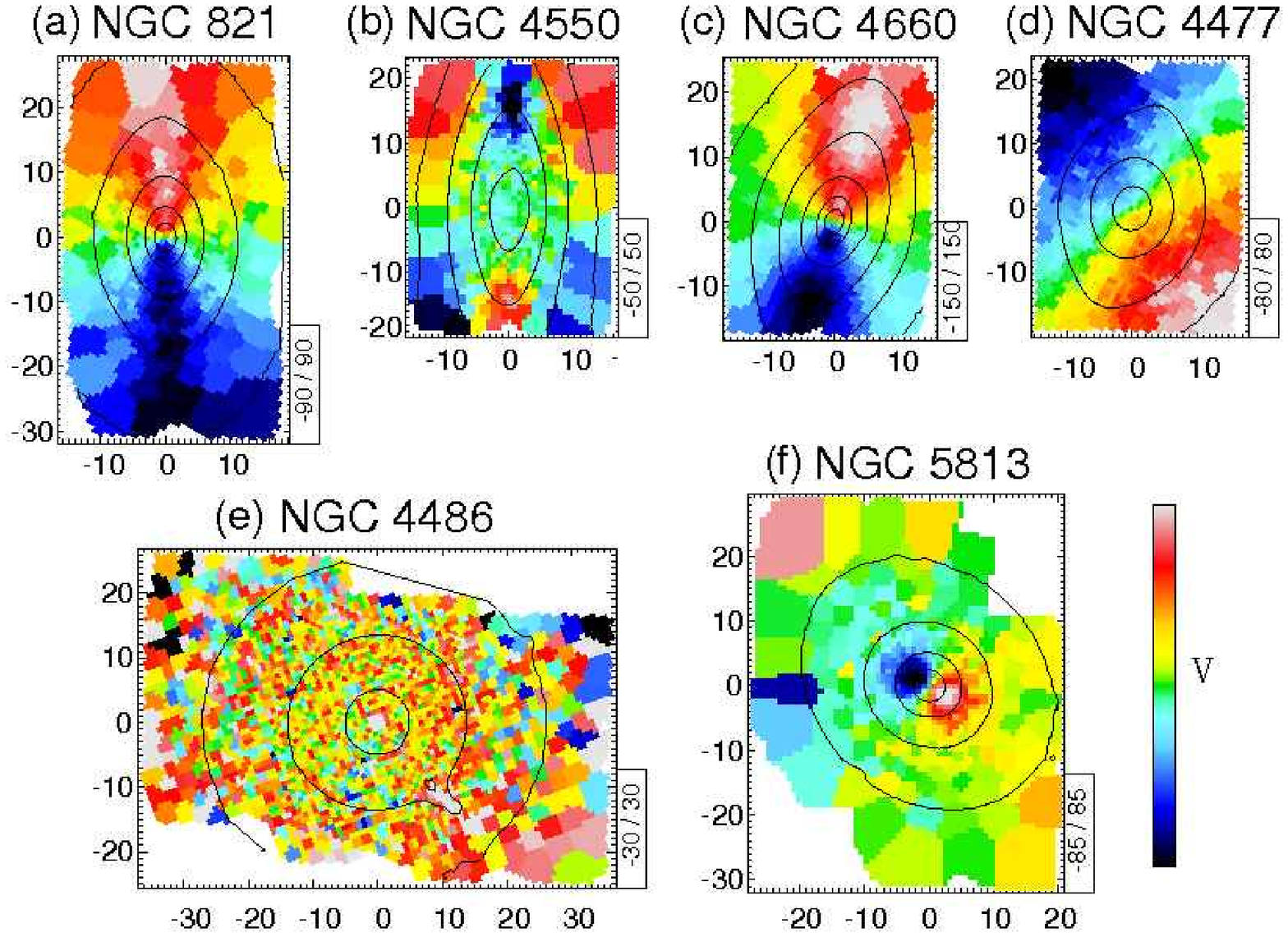}
\end{center}
\caption{Different velocity field morphologies observed with \sauron. (a) NGC\,821, (b) NGC\,4550, (c) NGC\,4660, (d) NGC\,4477, (e) NGC\,4486, (f) NGC\,5813. The maximum and minimum velocities (in \kms) plotted are indicated in the vertical tab. The data are spatially binned using the Voronoi tesselation method of \citet{cappellari03} to obtain a minimum signal-to-noise ratio of 60 per spectral resolution element, and isophotes are overlaid. (See Paper III for details)}
\label{fig:sauron_zoo}
\end{figure}

The stellar kinematics presented in Paper III exhibit a wealth of structure. Figure \ref{fig:sauron_zoo} shows examples of the kinds of different velocity field morphologies found in our sample. There are a large number of objects which show symmetric rotation around a single axis, consistent with a disk-like structure embedded in a hotter bulge-like component (e.g. NGC\,821, Figure \ref{fig:sauron_zoo}a). Figure \ref{fig:sauron_zoo}b shows a dramatic example of two counter-rotating disks with different scale heights in NGC\,4550, unambiguously determined from these data. Secondary central disks are also present in a number of objects (e.g. NGC\,4660, Figure \ref{fig:sauron_zoo}c), visible as a second peak in the velocity distribution near the center of the object. As well as aligned rotation, several objects show a rotation axis that is misaligned with the local photometric minor-axis, and which can vary in orientation across the field (e.g. NGC\,4477, Figure \ref{fig:sauron_zoo}d), indicating a non-axisymmetric system, such as a bar or triaxial figure. Some galaxies show negligible rotation (e.g. M87, Figure \ref{fig:sauron_zoo}e), of which most are giant elliptical galaxies. And several objects show KDCs, which clearly rotate around a different axis compared to the main body of the galaxy, and in some cases in the opposite direction (e.g. NGC \,5813, Figure \ref{fig:sauron_zoo}f).\looseness-2

Generalising these different kinematic morphologies, there are basically two types of behaviour found in the velocity maps of our sample galaxies: galaxies exhibiting regular and ordered rotation across the field, which is generally flat or increasing at the edge of our observations (typically 1~R$_e$) - these correspond to objects like those shown in Figure \ref{fig:sauron_zoo}a-d; and galaxies that show little rotation, except for a possible KDC, and which have decreasing or zero rotation at the edge of our field-of-view - these correspond to objects like those shown in Figure \ref{fig:sauron_zoo}e-f. We refer to these two types of object as `fast-rotators' and `slow-rotators' respectively. This separation of objects based on the morphology of their velocity fields does not correlate exactly with the classical division of `elliptical' and `lenticular'. More specifically, there are a number of fast-rotating objects classified as elliptical from their surface brightness morphology alone, even though dynamically they have more in common with lenticular objects \citep[see also][]{kormendy96}. 

\subsection{Kinematically Decoupled Components with \sauron}

We focus here on the KDCs found in our \sauron\/ observations. As described in the introduction, the presence of a KDC is often taken as evidence for a major galaxy merger, in which the core of one galaxy becomes embedded in the center of the merger remnant. Due to the random orientation of the merger, this scenario can naturally account for the generally misaligned orientation of the embedded sub-component. The KDCs found with \sauron\/ tend to reside in slow-rotating galaxies, where the outer parts are not rotating significantly. Such a hot, pressure-supported outer body may also be the natural consequence of a major merger event. In the hierarchical paradigm, larger systems form more recently, and so such mergers should have happened quite recently.

The alternative scenario for the formation of a KDC is to `grow' the KDC via star-formation {\em in situ}. To explain the misalignment of the KDC, the material from which the stars form must itself be kinematically misaligned with the rest of the galaxy. It is difficult to explain how this could occur for material with an internal origin, accumulated from the ejecta of stellar evolution. Therefore the gas would most likely have an external origin.

In both of these scenarios, the KDC would most likely be composed of a different stellar population to that of the main galaxy. In general, studies using both ground- and space-based imaging and spectroscopy have found little evidence to suggest that KDCs have very different star-formation histories from their host galaxies \citep[e.g.][but see \citealt{bender92}]{forbes96,carollo97,davies01}. This suggests that, rather than having formed by the most recent merger, KDCs (or at least the stars from which they are made) were formed at early epochs.

\begin{figure}
\begin{center}
\includegraphics[height=12cm, angle=0]{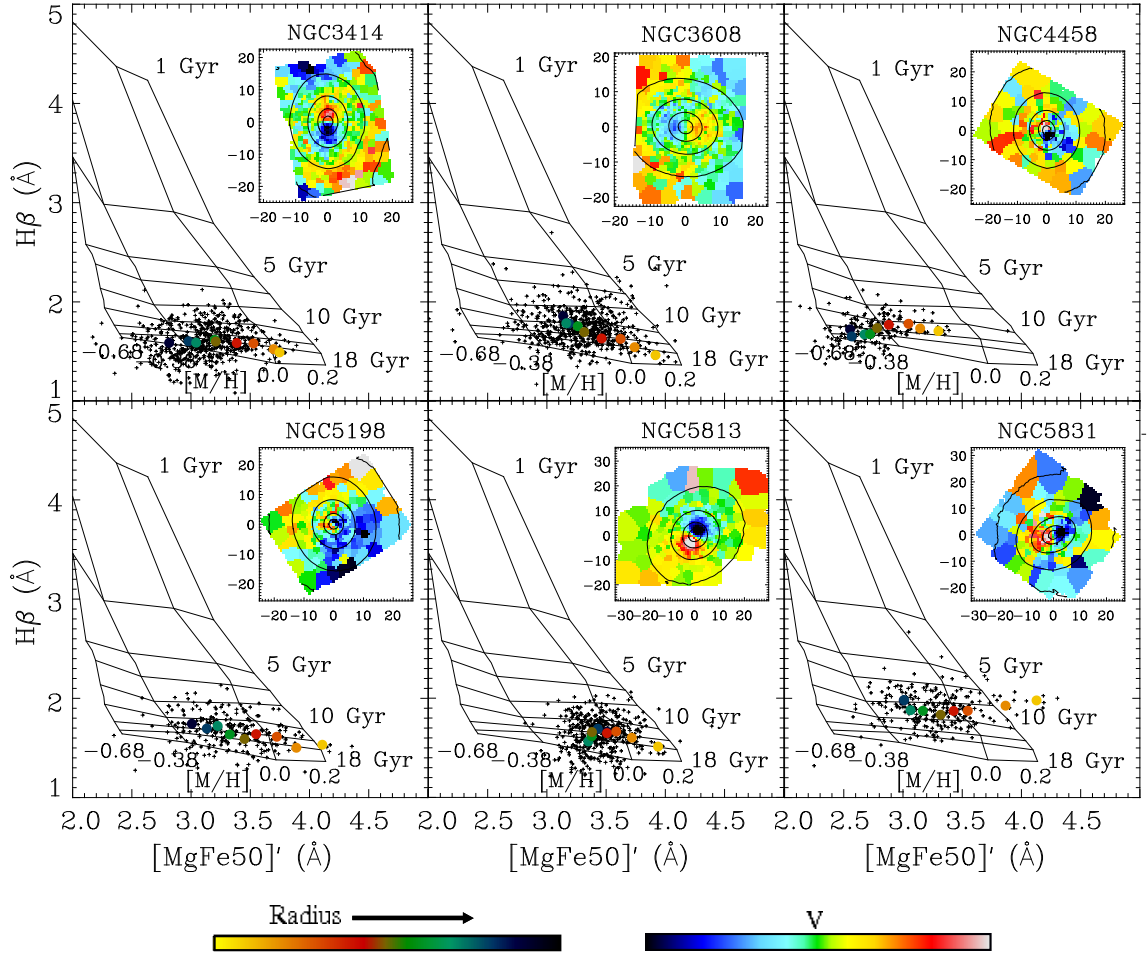}
\end{center}
\caption{Line-strength indices of six galaxies from the representative sample of 48 E/S0s presented in Paper III which show clear KDCs. Small symbols show the values measured in each (spatially binned) \sauron\/ aperture. Large symbols show values averaged along isophotes, with lighter symbols indicating smaller radii. In all cases, the populations are relatively old, and show a steady metallicity increase towards the centre with very little change in age between the outer body and KDC region. The inset images show the \sauron\/ velocity fields for reference, where the spatial scale is in arcseconds.}
\label{fig:sau_kdcs}
\end{figure}

Integral-field spectroscopy is the ideal tool for studying KDCs. Measuring the stellar kinematics and populations over a significant area of the galaxy provides an accurate picture of the KDC shape and orientation, and also allows structures to be directly associated by their spatial distribution. Figure \ref{fig:sau_kdcs} presents preliminary stellar population analyses of six galaxies from the \sauron\/ survey which exhibit clear and well-resolved KDCs. The velocity fields for these galaxies are given in the insert. The small symbols indicate the line-strength measurements obtained in every spatially-binned aperture of our data, typically involving a few hundred points. The large symbols show the same values averaged along isophotes. Lighter symbols imply smaller radii. Overplotted on the line-strengths is a grid of varying age (horizontal lines) and metallicity (vertical lines) from the stellar population models of \citet{vazdekis99}. This figure shows that, in all cases, the central KDC shows very little evidence of young stars. Indeed, these objects are predominantly old ($> 8$~Gyr), with remarkably small age gradients. All objects do show a smoothly increasing metallicity towards the center of the galaxy. Such metallicity gradients are common in early-type galaxies, however, and do not strongly distinguish these galaxies from non-KDC objects. We note that our parent sample of 48 objects is representative, not complete, and that examples of large KDCs with younger stellar populations may well exist, but are not found in our survey.

From this analysis, it appears that the KDCs in our sample are composed of old, evolved stellar populations, which, while exhibiting dramatically different kinematics from the rest of the galaxy, show little evidence to suggest a significantly different evolution to that of the main body of the galaxy. This is consistent with previous findings, and implies that these KDCs were formed early in the evolution of these galaxies. Although these objects appear kinematically peculiar, in a triaxial potential such orbital configurations may remain stable over a Hubble time. Over long timescales ($\sim 10$~Gyr), differences in stellar population age become less pronounced, making populations which differ in age by a few Gyr much more difficult to separate in older objects. All of this points to a picture where these KDCs are formed at early epochs, and then evolve quiescently until the current day. The specific formation mechanism, via either galaxy merging or `growth' from accreted material, is difficult to determine, as whatever differences exist between the star-formation histories of the KDC and its host galaxy are now diluted by the sands of time.

An alternative explanation may be that the {\em stars} in these systems formed at early epochs, but the KDC galaxies we observe have themselves {\em assembled} only recently via a dissipationless merger between two gas-free spheroidal systems. It has been shown that such `dry mergers' do occur, at least in cluster environments \citep{vdokkum99}, and that they may play a significant role in the latter stages of massive galaxy assembly \citep{khochfar03,bell05}. Distinguishing this scenario from an intrinsically early formation of KDC galaxies as we see them today is not trivial. An indicator of assembly at early epochs may be the presence of old, disk-like KDCs, since any pre-existing stellar disk would be significantly dynamically heated during a recent major merger. Determining the dynamical structure of the KDCs themselves is beyond the scope of this contribution, but will be considered in forthcoming papers of the \sauron\/ series. Also, in the dry-merger scenario, the remnant KDC is formed by the more compact and tightly-bound progenitor. It is difficult to reconcile why the progenitors, while strongly differing in density, should have such similar stellar populations.

\section{\oasis\/ Observations of Galaxy Centres}

The central regions of early-type galaxies are crucial to our understanding of galaxy formation and evolution. Since the advent of the Hubble Space Telescope (HST), high spatial-resolution studies of these galaxy nuclei have made key discoveries in their structure, which are apparently linked to their large-scale properties. Spectroscopic studies with HST have given the strongest evidence for the existence of super-massive black holes in the centres of galaxies other than our own \citep[e.g.][and references therein]{kormendy95}. These black holes are thought to be ubiquitous, at least amongst spheroidal systems, and have masses which are closely linked to the mass of the host system \citep[e.g.][]{ferrarese00,gebhardt00}. Galaxy formation theories suggest that the co-evolution of black holes and their host galaxies is a fundamental aspect of how galaxies form, where accretion of material onto the black hole controls star-formation efficiency via feedback mechanisms \citep[e.g.][]{silk98,granato04,binney04,springel05}.\looseness-2

In addition, the central light profiles of early-type galaxies show differing gradients at HST resolution, ranging from a continuously rising or steepening profile (`cusp'), to a flat profile (`core') with a distinct break in slope from the outer parts \citep{jaffe94,lauer95,lauer05}. These light profile types correlate with other global galaxy parameters, such as disky or boxy isophotes, and the degree of rotational support \citep{faber97}. The role of the central black hole is thought to be important in determining the shape of the central light profile. Cusp nuclei form through the build-up of material in the deep potential well of the black hole environment, which then forms stars giving a central peak in the light profile. Core nuclei, on the other hand, are thought to form through `scouring' of the central regions by a coalescing black hole binary. This binary forms as the result of a galaxy merger, where the central black holes of the merging galaxies quickly fall to the center of the potential. The binary hardens by losing angular momentum to the nearby stars, which are ejected from the core region. The result is a deficit of stars in the environment of the central black hole, giving a flat light profile \citep[e.g.][]{quinlan97,merritt01}.\looseness-2

For our survey, the spatial sampling of the \sauron\/ spectrograph was set to 0\farcsec94 $\times$ 0\farcsec94 per spatial element. This gives maximum field of view (33\arcsec\/ $\times$ 43\arcsec), but often undersamples the typical seeing at La Palma (0\farcsec7--0\farcsec8 FWHM). For the main body of the galaxies, this is not important, since the properties of early-type galaxies generally vary smoothly on scales larger than the seeing. In the central regions of these objects, however, the light profile can rise steeply, becoming unresolved even in the best seeing. In these regions, the coarse spatial sampling of \sauron\/ becomes a problem, resulting in poorly resolved measurements of the galaxy centres. For this reason, we are conducting a campaign of follow-up observations of the centres of the \sauron\/ survey E and S0 galaxies, using high spatial-resolution integral-field spectroscopy. The first steps in this project have been taken with the \oasis\/ spectrograph during its operation at the Canada-France-Hawaii Telescope (CFHT), on Hawaii. In July 2003, \oasis\/ was moved to the WHT to operate behind the NAOMI Adaptive Optics (AO) system, which is described by Benn et al. in these proceedings.\footnote{The capabilities of \oasis\/ at the WHT are described in \citet{mcdermid04b}}

In total, 28 of the 48 \sauron\/ E and S0 galaxies have been observed, using a wavelength region and spectral resolution comparable to that of \sauron. The observations were seeing-limited, due to a lack of suitable natural guide stars. However, the spatial sampling of \oasis\/ was set to 0\farcsec27 $\times$ 0\farcsec27 per spatial element, thus adequately sampling even the best conditions on Mauna Kea. The median spatial-resolution of the \oasis\/ observations is almost a factor two better than that of our \sauron\/ observations of the same objects. A full description of the sample, observations and data reduction, as well as measured parameter maps is given in \citet{ mcdermid05}.

\subsection{Comparing \sauron\/ and \oasis}
\label{sec:oas_sau_comp}

\begin{figure}
\begin{center}
\includegraphics[width=14cm, angle=0]{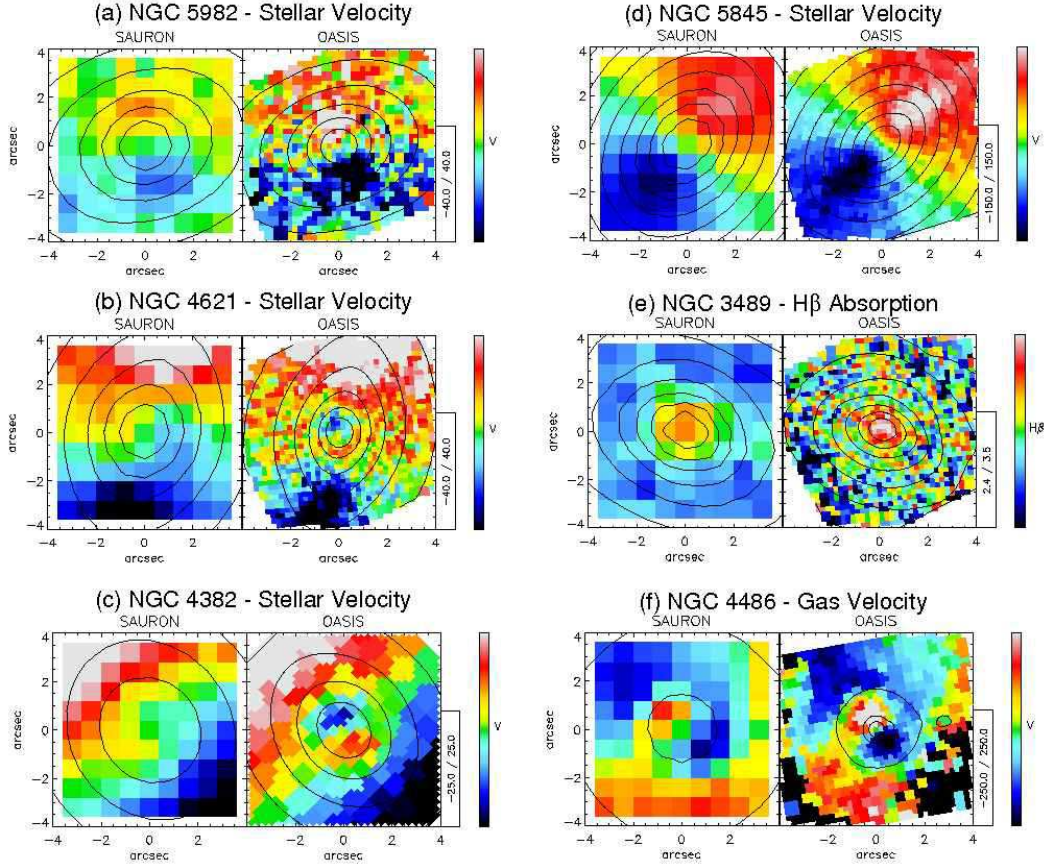}
\end{center}
\caption{Comparison of \sauron\/ ({\em left}) and \oasis\/ ({\em right}) observations of six early-type galaxy centres. The field sizes are the same between the pairs of panels, as are the intensity scales. Isophotes from the integrated spectra are overlaid. The maximum and minimum levels are given in the inserted tab, where units are \kms\/ for velocity fields and \AA\/ for line-strength indices. (a) Stellar velocity field of NGC\,5982. (b) Stellar velocity field of NGC\,4621. (c) Stellar velocity field of NGC\,4382. (d) Stellar velocity field of NGC\,5845. (e) H$\beta$ absorption map of NGC\,3489. (f) Gas velocity field of NGC\,4486.}
\label{fig:sau_oas_comp}
\end{figure}

It is instructive to explore what additional information we obtain by observing the \sauron\/ objects with higher spatial-resolution by considering some illustrative examples. Figure \ref{fig:sau_oas_comp}a shows the stellar velocity field of NGC\,5982 measured with \sauron\/ and \oasis. The center appears misaligned in the \sauron\/ observations, indicating a probable decoupled component or kinematic twist. This is confirmed by \oasis, showing that the misaligned component actually rotates rapidly around the major axis of the galaxy, indicating a prolate core.\looseness-2

Figure \ref{fig:sau_oas_comp}b shows the stellar velocity field of NGC\,4621. The `kink' in the zero-velocity contour observed with \sauron\/ is resolved into a clearly counter-rotating core, which was previously found using \oasis\/ with AO \citep{wernli02}. Similarly, Figure \ref{fig:sau_oas_comp}c shows that NGC\,4382 contains a clear decoupled core when observed with \oasis. Figure \ref{fig:sau_oas_comp}d further shows that it is not only misaligned and counter-rotating central components which are measured with far greater accuracy using \oasis. NGC\,5845 harbours an edge-on, thin central disk, which is clearly visible in HST imaging. The \sauron\/ observations indicate the presence of a central rotating component, although the amplitude of rotation does not conclusively indicate a dynamically cold disc. The \oasis\/ observations, on the other hand, beautifully resolve this disk component, showing rapid rotation tightly confined to the major-axis of the galaxy.

Other spectral properties show improvement when measured with \oasis. Figure \ref{fig:sau_oas_comp}e compares the H$\beta$ absorption index observed with \sauron\/ and \oasis\/ for NGC\,3489, which shows strong H$\beta$ absorption in the center. The improved image-quality of the \oasis\/ observations shows that the strength of absorption in the center is increasing within the resolution limit of \sauron. This makes a significant difference to the luminosity-weighted age determined in the central lenslets. Likewise, the kinematics of ionised gas can show complex distributions around galaxy nuclei. Figure \ref{fig:sau_oas_comp}f shows the gas velocity field for NGC\,4486 (M87). \sauron\/ shows a stream of gas spiraling to the center of the galaxy, connecting smoothly to a component rotating around the center. The \oasis\/ observations show that this central component rotates much more rapidly than the surrounding gas, and in fact is probably related to the Keplerian disk of material orbiting the central super-massive black hole \citep{ford94}.\looseness-2

\subsection{Stellar Kinematics of Young Nuclei}

Our \sauron\/ observations show that in the majority of galaxies exhibiting young stellar populations, the young stars are strongly concentrated towards the galaxy center, rather than distributed evenly throughout the object. This is in contrast to the distribution of ionised gas, which can exist in regular, large-scale structures extending beyond an effective radius (Paper V, see also Falc\'on-Barroso et al. in this proceedings). It would seem that only in the central regions, deep in the galaxy potential well, are the conditions suitable to trigger the conversion of the gas reservoir into stars. What are the causes and implications of this star-formation? What is the relationship between the star-formation and the stellar potential? What is the significance of this star-formation to the global evolution of the system?

\begin{figure}
\begin{center}
\includegraphics[width=14cm, angle=0]{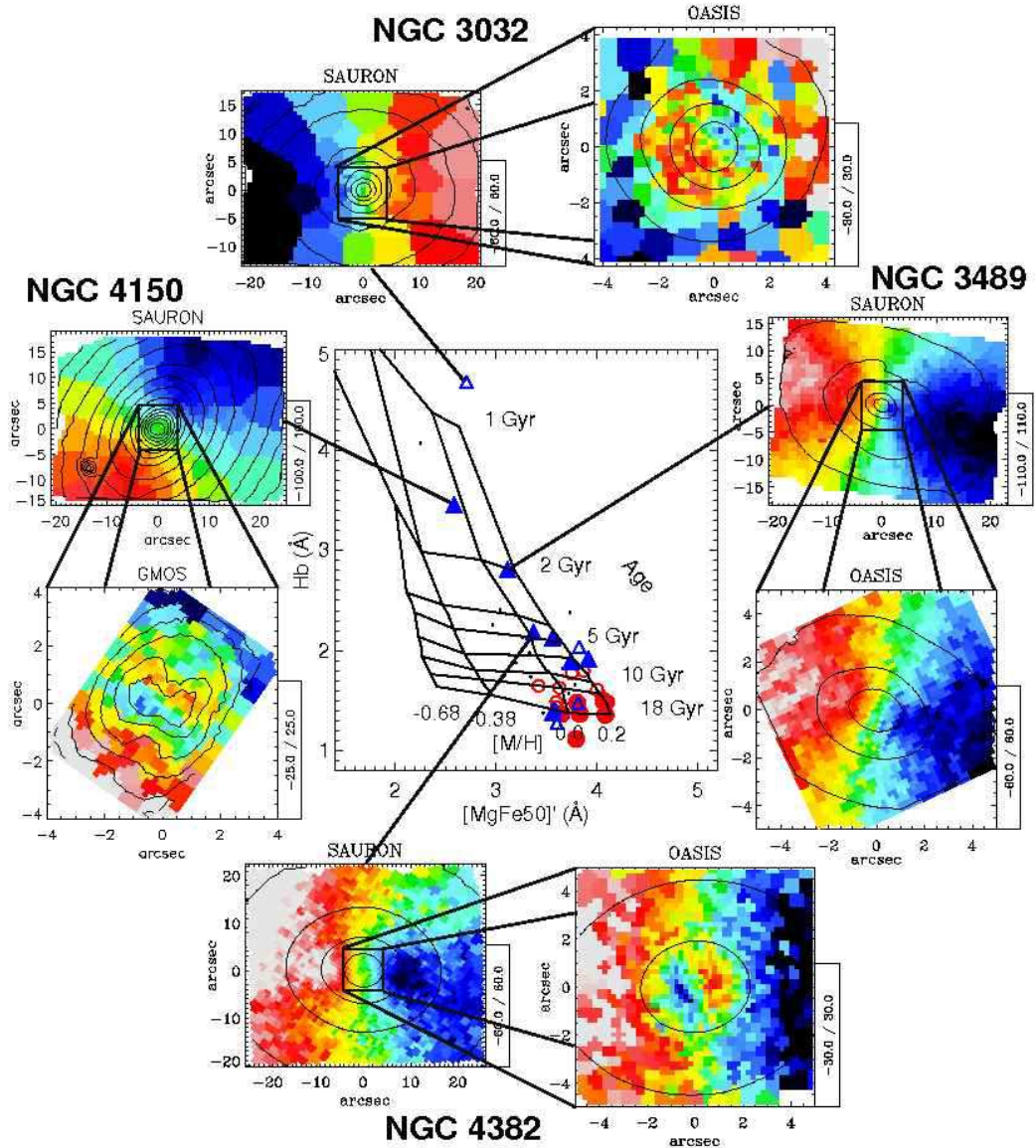}
\end{center}
\caption{{\em Central panel:} Line-strength indices measured within an R$_e/8$ circular aperture on our \sauron\/ data from Paper VI, overplotted with model stellar population predictions from \citet{vazdekis99}. Large symbols show galaxies observed in our \oasis\/ sub-sample. Small points show the remaining objects in the full \sauron\/ E/S0 sample. Triangles and circles indicate E and S0 galaxies respectively. Open symbols show `field' objects, and filled symbols `cluster' objects, as described in Paper II. {\em Insert plots:} \sauron\/ stellar velocity fields of the indicated galaxies, with a secondary `zoom' on the central regions showing our high spatial-resolution \oasis\/ measurements, except in the case of NGC\,4150, where we present data from the \gmos\/ integral-field spectrograph.}
\label{fig:oas_kdcs}
\vspace{3cm}
\end{figure}

Our \oasis\/ observations are the ideal tool to investigate these questions. The central panel of Figure \ref{fig:oas_kdcs} shows the distribution of aperture line-strength values, measured within an R$_e/8$ circular aperture with our \sauron\/ data. The large symbols indicate objects also observed with \oasis. Of the \oasis\/ sub-sample, four objects show relatively strong H$\beta$ absorption, with luminosity-weighted age $<5$~Gyr. The surrounding panels show, for each of these four objects, the \sauron\/ stellar velocity field, with a central insert showing the stellar velocity field observed at higher spatial-resolution with either \oasis, or in the case of NGC\,4150, with \gmos\/ on Gemini-North.

We find that three of these four galaxies (NGC\,3032, NGC\,4150, and NGC\,4382) with strong H$\beta$ absorption also exhibit a clearly decoupled component in their stellar kinematics. The KDCs in NGC\,3032 and NGC\,4382 have been discovered from our \oasis\/ observations \citep{mcdermid05}. The fourth object, NGC\,3489, does not exhibit the clear counter-rotation that the other objects show. Closer inspection of the velocity field does, however, indicate that there is a central fast-rotating sub-component. The curve of zero velocity also shows some evidence of a (spatially resolved) mild twist across the galaxy center, indicating that there may be some departure from regular axisymmetric rotation in the central parts.

\begin{figure}
\begin{center}
\includegraphics[width=14cm, angle=0]{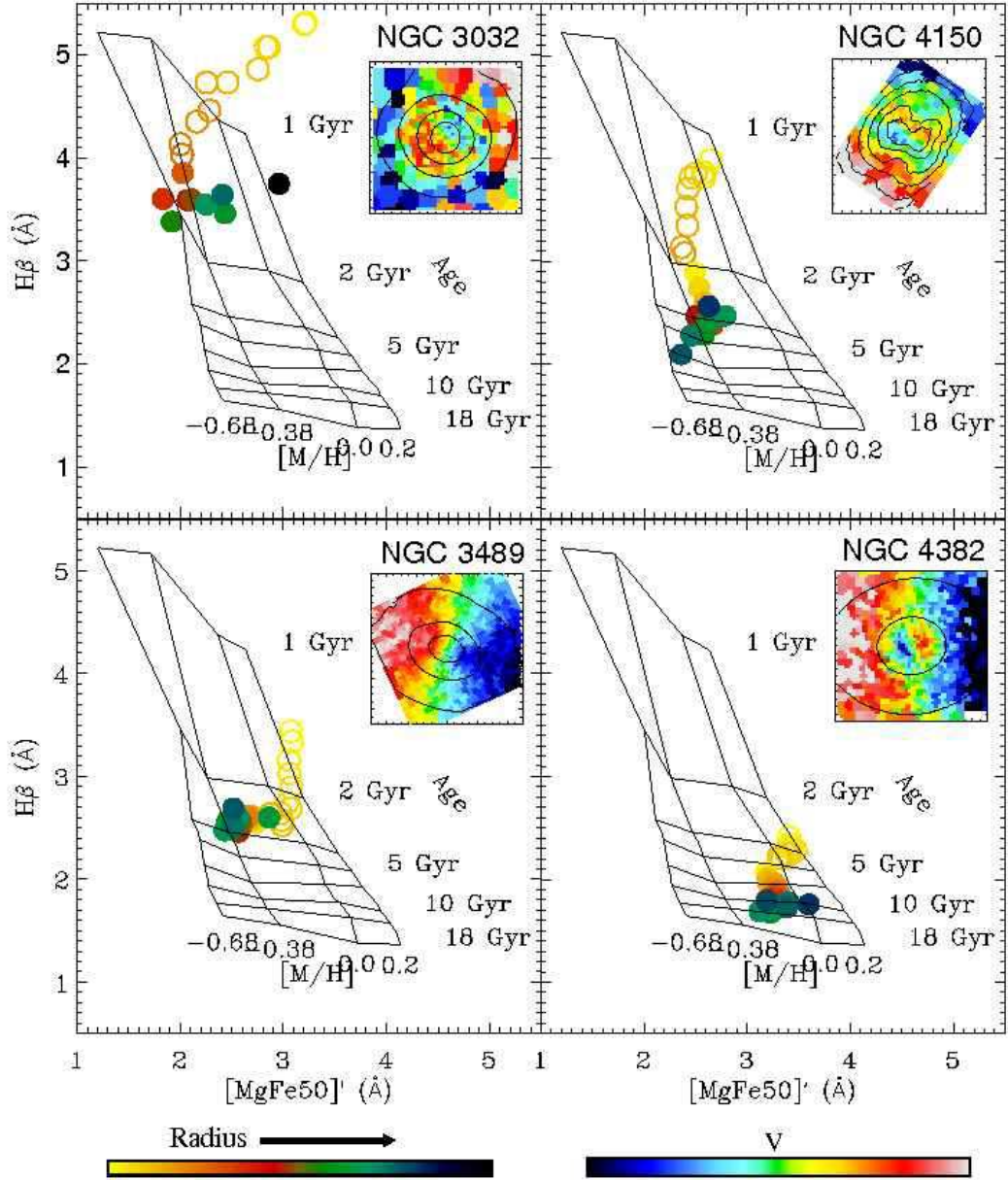}
\end{center}
\caption{Similar to Figure \ref{fig:sau_kdcs}, but for the four galaxies with central luminosity-weighted age $<5$~Gyr highlighted in Figure \ref{fig:oas_kdcs}. \oasis\/ and \gmos\/ measurements are included (open symbols) in addition to the \sauron\/ data (filled symbols), of which the central 3\arcsec\/ are excluded to avoid PSF effects. All four objects show a sharp decrease in luminosity-weighted age within the central few arcseconds, coincident with the KDC component in the high spatial-resolution data.\looseness-2}
\label{fig:oasis_kdcs_pops}
\vspace{3cm}
\end{figure}

Figure \ref{fig:oasis_kdcs_pops} presents the stellar population parameters, comparable to Figure \ref{fig:sau_kdcs}, but with the addition of our high spatial-resolution values inside the central 3\arcsec, shown as open symbols. This demonstrates that the central regions, where the decoupled components are located, show a significant age gradient towards younger populations. \sauron\/ data points inside 3\arcsec\/ are excluded here, as they may be biased by PSF effects.

\subsection{Stellar Populations of Kinematically Decoupled Components}

Comparing Figures \ref{fig:oasis_kdcs_pops} and \ref{fig:sau_kdcs}, it seems that there are basically two types of KDC: apparently large KDCs, which have populations indistinguishable from their host galaxy, and which exist in galaxies that tend to show little ordered rotation; and small KDCs, which are composed of young populations, and are found in galaxies that exhibit significant rotation. This was found, however, by considering only the four youngest objects in our \oasis\/ sub-sample. Here we explore the relationship, if any, between the intrinsic size of KDCs found in our sample (either from \sauron\/ or \oasis\/ data) and the stellar population within the KDC. We only consider KDCs which are unambiguously distinct from the rest of the galaxy, showing significantly misaligned or counter-rotating stellar motions (for example, NGC \,3489 shown above is not included here). This results in a list of 13 objects. To the 9 KDCs we have already considered in the above discussion, we add four objects: NGC\,4621 and NGC \,5982 - these objects appear to have KDCs from our \oasis\/ observations (see section \ref{sec:oas_sau_comp}); and NGC\,7332 and NGC\,7457 - although not observed with \oasis, these galaxies show small (unresolved) but distinct central twists in their \sauron\/ velocity fields \citep[Paper III, see also][]{falcon04}.

First we must quantify the spatial size of the KDC. From the map of mean stellar velocity, it is possible to determine where the contribution of the KDC to the line-of-sight velocity distribution becomes comparable to that of the outer body of the galaxy. At this point, the mean net rotation shows a minimum, and in some cases reaches zero, isolating the KDC from the rest of the galaxy. This point can be used to make an estimate of the KDC size, although there may still be an appreciable fraction of the KDC beyond this radius. For KDCs larger than the seeing FWHM, this will generally underestimate the size. For KDCs that appear similar in size to the seeing FWHM, this estimate may be too large. With this in mind, we use the largest diameter across the center of the galaxy which connects the zero velocity curve on opposite sides of the KDC.\looseness-2

\begin{figure}
\begin{center}
\includegraphics[width=12cm, angle=0]{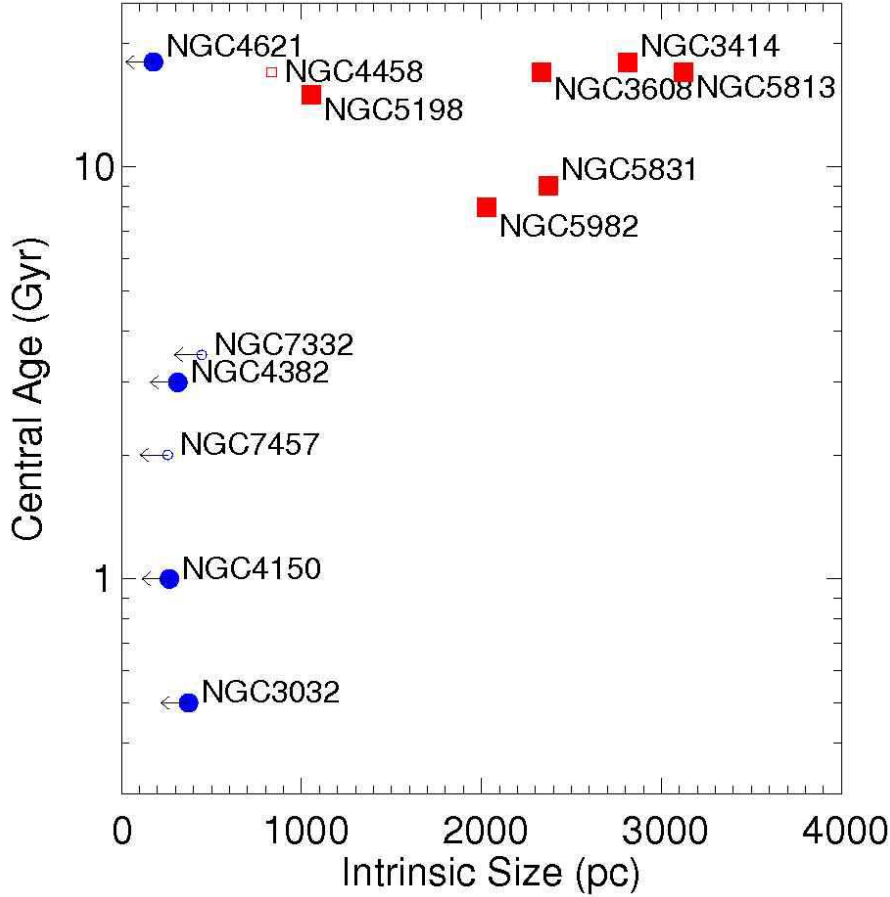}
\end{center}
\caption{Luminosity-weighted age of the central aperture plotted as a function of KDC intrinsic diameter. Ages are computed from the stellar population models of \protect\citet{thomas03}, and include element abundance ratios as a free parameter. Circlular symbols denote fast-rotating galaxies; squares denote slow-rotating galaxies. Arrows denote cases where the apparent diameter of the KDC is less than four times the seeing FWHM, implying that we only obtain an upper limit. Filled symbols have both \sauron\/ and \oasis\/ or \gmos\/ observations. Open symbols have only \sauron\/ data.}
\label{fig:kdc_size_age}
\end{figure}

We convert this apparent diameter into an intrinsic diameter using distances from surface brightness fluctuation studies \citep{tonry01}. Figure \ref{fig:kdc_size_age} plots the estimated intrinsic diameter of each KDC against the luminosity-weighted age derived from line-indices measured within the central pixel of the highest spatial-resolution data available (either \sauron, \oasis\/ or \gmos). From this plot, it can be seen that the intrinsically smallest KDCs (less than 400pc in diameter) also tend to exhibit the youngest populations. KDCs of around 1~kpc or more in diameter are all older than 8~Gyr.

There are a number of implications to these findings. In the initial discussion, we mentioned that KDCs found in early-type galaxies have often been used as evidence of recent merging activity. We also described how the spread in apparent ages of early-type galaxies has supported the scenario where these objects are amongst the last to form, as a hierarchical sequence of galaxy mergers. However, inspection of the stellar populations of `classical' KDC galaxies show that the KDC populations are coeval with the rest of the galaxy, and exhibit no peculiar properties to distinguish KDC galaxies from ordinary galaxies. We therefore cannot directly associate the {\em dynamical} evidence for recent merging and the {\em population} evidence for such merging by considering these objects. From our high spatial-resolution integral field spectroscopy of the subset of \sauron\/ survey galaxies, however, we have found that the youngest galaxies do exhibit KDCs which can be directly associated with the young population.

A tempting, but naive interpretation of this would be that the case of the missing young KDCs has been solved, and that these sub-components show both stellar dynamical and population evidence for recent merging. However, comparing the properties of the host galaxies reveals that the 100~pc-scale KDCs are not directly comparable to their kpc-scale counterparts. The different plotting symbols in Figure \ref{fig:kdc_size_age} denote the separation of objects into fast-rotators and slow-rotators, as discussed in section \ref{sec:sau_zoo}. Here we find a strong connection between the large, old KDCs and non-rotating galaxies, and the small, generally young KDCs and fast-rotating galaxies. The global properties of these objects, therefore, are rather different.

The fraction of the galaxy's mass contained in the KDC is also rather different between the two types of object. The young KDCs generally extend to only $\lsim 0.1$R$_e$, whereas the older KDCs can extend up to $\sim 0.4$R$_e$. Although this is only a rough indication of the actual mass fraction of the KDC, we should remember that the small KDCs have also young populations with low M/L values, thus contributing significantly to the light, but relatively little in mass.\looseness-2

We therefore conclude that these two manifestations of KDCs are distinct, both in origin and destiny. On one hand, large, old KDCs are consistent with having formed through major merging (having a large mass fraction, and residing in dynamically hot triaxial systems), being either dissipative at early epochs, followed by largely quiescent evolution thereafter; or dissipationless at more recent times. On the other hand, compact, young KDCs appear to be consistent with more modest dissipational accretion events, residing in fast-rotating disk-like galaxies containing plenty of star-forming material. In time, this young population will evolve and diminish in brightness, and the KDC may reduce in apparent size as it becomes less distinct above the background of the rest of the galaxy. The precise connection between the dynamical decoupling of the stars and the recent or ongoing star-formation in these young KDCs requires further investigation, and it is not yet clear whether the decoupled stars are formed from seed material which is dynamically distinct from the rest of the galaxy; or whether pre-existing decoupled stars themselves trigger star-formation as material enters the KDC potential. The dynamical structure of the young KDCs is also uncertain, and difficult to determine given their embedded nature and the limited spatial resolution of our data. However, the centrally concentrated nature of the young stars suggests an efficient transfer of material to the galaxy center. Furthermore, although close to being aligned, most (if not all) of the young KDCs exhibit some offset in their rotation axis to that of the rest of the galaxy. This points towards the involvement of asymmetry in the central potential, and possibly the presence of a nuclear bar. In any case, these objects seem to be driving the large spread in age observed for early-type galaxies, and the existence of a small KDC seems closely linked to the associated star-formation.

\section{Conclusions}

The study of nearby early-type galaxies is crucial to our understanding of galaxy formation and evolution, and sets key constraints for complementary high-redshift studies. The advent of integral-field spectroscopy is heralding a new era in our observational understanding of these objects. In this contribution, we have reviewed some of the key observational constraints on how early-type galaxies form and evolve, looking specifically at the connection between the apparent spread in luminosity-weighted age of these galaxies and the existence of kinematically decoupled sub-components, in the context of galaxy formation through hierarchical merging. From our wide-field \sauron\/ integral-field spectroscopy, we have shown that the clearly detected KDCs in our sample exhibit old stellar populations, and are coeval with their host galaxies. This strongly suggests that these KDCs are not the product of recent merging, but instead were formed at early epochs (either through merging or from accreted material which subsequently formed stars), and that no such significant evolutionary event has occurred in these objects since. We note, however, that formation via recent `dry' merging cannot be strongly excluded.

We have introduced a series of follow-up observations on the central regions of the \sauron\/ E and S0 galaxies, obtained at higher spatial-resolution using the \oasis\/ integral-field spectrograph. These observations reveal a wealth of structure in the central regions of these galaxies which is poorly resolved with \sauron. In particular, we find that the youngest galaxies in our current \oasis\/ sub-sample show clear evidence of harbouring compact central KDCs, which were previously unresolved by \sauron\/ or other studies. The apparent size of these KDCs are consistent with the spatial extent of the young populations, implying that the two phenomena are directly linked.

Estimating the intrinsic size of the various KDCs found in our full sample, either with \sauron\/ or \oasis, we find that, whilst small (100~pc-scale) KDCs can exhibit a large range of ages, larger (kpc-scale) KDCs are generally old ($> 8$~Gyr). We find that there is a connection between these two types of KDC and their host galaxy, such that young, compact KDCs are found in fast-rotating objects, whereas old, large KDCs are found in slowly-rotating systems. We conclude that these objects have distinct formation histories, and that these young KDCs are not strong evidence for the formation of early-type galaxies via recent merging. Further investigation will reveal how the dynamical decoupling of the central stars is related to recent or ongoing star-formation in these objects. 

We have extended this sub-sample of the 48 \sauron\/ E/S0 by obtaining similar quality data for the lowest velocity dispersion objects using the \gmos\/ spectrographs on Gemini North and South, and hope to observe the remaining objects in due course with \oasis\/ at the WHT. This collection of data will be used to complement ongoing studies based on the \sauron\/ project, including using them to constrain general dynamical models. With these high-spatial-resolution integral-field measurements, and including HST spectroscopy where available, we will place strong constraints on the central orbital structure, in order to probe the formation of central super-massive black holes (Cappellari \& McDermid, 2005). These seeing-limited observations will form the foundation of future studies that will make use of the enhanced spatial-resolution offered by adaptive optics facilities. Development of laser guide star systems, such as GLAS (see the contribution by Rutten in these proceedings) will greatly expand the sky coverage for AO-fed integral-field spectrographs, such as \oasis\/ (WHT), \sinfoni\/ (VLT), \osiris\/ (Keck) and \nifs\/ (Gemini), allowing systematic study of large, representative samples of nearby nuclei at high spatial resolution.\looseness-2
\newpage
{\bf ACKNOWLEDGMENTS}

RMcD would like to thank R.~Bacon for presenting this talk on his behalf - Baby McDermid was born one week later. It is a pleasure to thank the ING staff, both for the organisation of a fruitful and timely workshop, and for their continuuing support of the \sauron\/ project and instrument.

{\small Based in part on observations obtained at the Gemini Observatory, which is operated by the Association of Universities for Research in Astronomy, Inc., under a cooperative agreement with the NSF on behalf of the Gemini partnership: the National Science Foundation (United States), the Particle Physics and Astronomy Research Council (United Kingdom), the National Research Council (Canada), CONICYT (Chile), the Australian Research Council (Australia), CNPq (Brazil) and CONICET (Argentina).}

\end{document}